\documentclass[onecolumn,amsmath,amssymb,12pt,nofootinbib]{revtex4}
\usepackage{graphicx}

\newcommand{\nn}{\nonumber}
\newcommand{\be}{\begin{equation}}
\newcommand{\ee}{\end{equation}}
\newcommand{\bea}{\begin{eqnarray}}
\newcommand{\eea}{\end{eqnarray}}

\def\a{\alpha}

\def\d{\delta}

\def\s{\sigma}

\def\e{\epsilon}

\def\f{\phi}

\def\k{\kappa}

\def\th{\theta}

\def\s{\sigma}

\def\t{\tau}

\def\z{\zeta}

\def\pt{\partial}

\def\ie{{\it i.e. }}

\begin{document}

\title{Diversity in the Phoenix Universe}

\author{
Jean-Luc Lehners
}

\affiliation{\small Max-Planck-Institute for Gravitational Physics (Albert-Einstein-Institute) \\
      D-14476 Potsdam/Golm, Germany }

\begin{abstract}
It has recently been argued by Copeland {\it et al.} that in eleven dimensions two orbifold planes can collide and bounce in a regular way, even when the bulk metric is perturbed away from Milne spacetime to a Kasner solution. In this paper, we point out that as a consequence the global ``phoenix'' structure of the cyclic universe is significantly enriched. Spatially separated regions, with different density fluctuation amplitudes as well as different non-gaussian characteristics, are all physically realized. Those regions containing by far the most structure are specified by a fluctuation amplitude of $Q \sim 10^{-4.5}$ and local non-gaussianity parameters $f_{NL} \sim {\cal O}(\pm 10)$ and $g_{NL} \sim {\cal O}(-10^3),$ in agreement with current observations.

\end{abstract}

\maketitle

\section{Introduction}

In cosmology, it is tempting to declare that we should only ever think about what happened in our past lightcone, as everything outside of it is inaccessible to observations. However, several well-motivated models of the early universe predict the existence of regions of the universe, with a whole range of physical properties, that are far separated in space and/or in time from our observable region of the universe. Hence it may be important to understand the characteristics of these far removed regions, at least on a theoretical level. It certainly seems of interest to explore all the consequences of a given cosmological model, whether these appear to be immediately observable or not; this is crucial in assessing the logic and internal consistency of the model in question, and hence crucial in determining whether a given cosmological model satisfies us intellectually. Such an understanding has concrete implications, as the perception of which properties of our region of the universe we consider to be determined by mathematical necessity rather than historical accident guides us in devising future research avenues.

The question of additional universes is particularly acute for eternal inflation \cite{Guth:2007ng}, since the number and variety of universes are both infinite\footnote{We will occasionally use the expressions ``universe'' and ``region of the universe'' interchangeably, when this does not seem confusing.}. Hence all possible universes are produced an infinite number of times, so that, taken by itself, the model of eternal inflation is not predictive. In order to make it predictive one needs to add a measure, which regulates the infinities in question and leads to definite predictions. Unfortunately it turns out that all predictions depend sensitively on the measure/regulator used \cite{Freivogel:2011eg} -- in other words, unlike the situation in QED for example, all predictions are regulator-dependent. Moreover, some of the simplest measures turn out to be in blatant conflict with observations. Hence we are certainly missing something important, and one may even conjecture that eternal inflation, as currently envisaged, is unphysical and does not occur.

In this paper, we will examine the question of predictivity in an alternative cosmological model, namely the cyclic theory of the universe proposed by Steinhardt and Turok \cite{Steinhardt:2001st}. In doing so, we will extend existing studies of the global structure of the cyclic universe, previously described as the {\it phoenix universe} \cite{Lehners:2008qe,Lehners:2009eg}. Building on recent results by Copeland, Niz and Turok regarding brane collisions \cite{Copeland:2010yr}, we will show that the cyclic universe also leads to vastly separated regions with distinct physical properties. Hence one may be worried that a similar ambiguity as the one inherent to eternal inflation will come to plague the model. However, as we will describe, the cyclic model predicts strong correlations between a large number of physical observables in a given region, such as its spatial flatness, the amount of structure it contains and the non-gaussian statistics of its density perturbations. These correlations are entirely determined by the dynamics of the model, and thus they lead to clear, testable, predictions. For the specific model described below, it turns out that those regions that are flat and contain the most structure are characterized by a fluctuation amplitude of $Q \sim 10^{-4.5}$ and local non-gaussianity parameters $f_{NL} \sim {\cal O}(\pm 10)$ and $g_{NL} \sim {\cal O}(-10^3).$ Thus the model is both in agreement with current observations, and falsifiable by upcoming CMB experiments.

Before proceeding, we should clarify that the cyclic model contains a number of open questions. The most important such open question has to do with the bounce that connects the ekpyrotic contracting phase with the phase of expansion. This big crunch/big bang transition has been modeled by a collision of orbifold branes in M-theory \cite{Khoury:2001wf}. To date, it has only been possible to analyze such a collision at the semi-classical level \cite{Turok:2004gb,Copeland:2010yr}, and not at the full quantum level. In the present work, we take the semi-classical results seriously, and explore their consequences. Thus, the present work is only applicable insofar as the semi-classical calculations accurately reflect the physics of the bounce. But, to the extent that our analysis yields promising results, it also motivates further work on the quantum treatment of the crunch/bang transition.

The plan of the paper is as follows: we first provide a brief review of the cyclic universe, focussing on those aspects that are most relevant to the present paper. Section \ref{SectionPhoenix} then discusses the global structure of the cyclic universe, reviewing the picture as it was understood up to now, and extending it by using the results of recent semi-classical calculations of colliding orbifolds. The new global phoenix structure of the cyclic universe, together with its implications, is discussed. Conclusions and further directions are to be found in the discussions section.

\section{The Cyclic Universe} \label{SectionCyclic}

\begin{figure}[t]
\begin{center}
\includegraphics[width=0.75\textwidth]{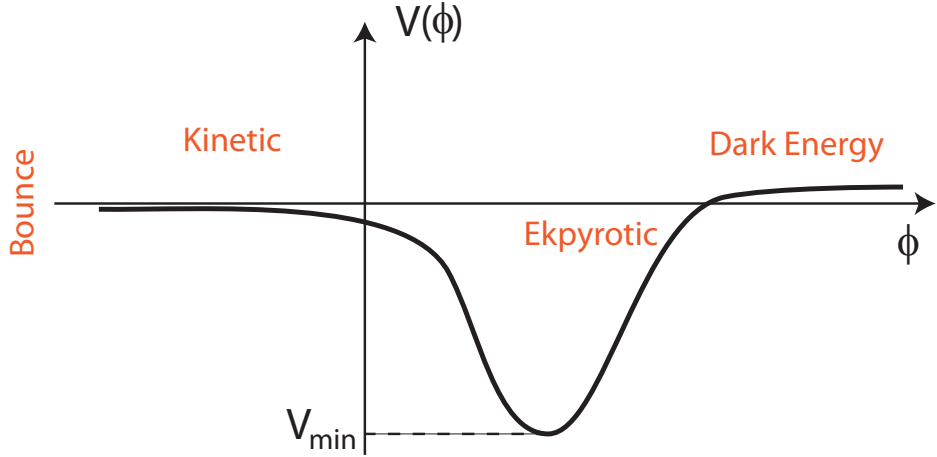}
\caption{\label{figurePotential} {\small The cyclic universe potential and its main phases of evolution.
}}
\end{center}
\end{figure}

In the cyclic universe \cite{Steinhardt:2001st,Lehners:2008vx,Lehners:2011kr}, periods of expansion alternate with periods of contraction. During the expanding phase, the energy density of the universe is successively dominated by radiation, dark matter and dark energy. During this phase, the universe grows by a huge factor, of the order of $exp(60+N_{de}),$ where $N_{de}$ denotes the number of e-folds of dark energy domination. Towards the end of the expanding phase, the scalar field responsible for dark energy starts to roll faster down its potential, until eventually the potential becomes negative - see Fig. \ref{figurePotential}. Shortly thereafter, the universe reverts from expansion to contraction, and an ekpyrotic phase ensues. The ekpyrotic phase is characterized by a very stiff equation of state (\ie the pressure is much larger than the energy density during this phase), which has the consequence of suppressing both the average curvature and any anisotropies in the curvature of the universe, thus keeping/rendering the universe flat and isotropic \cite{Erickson:2003zm}. During this phase, the scale factor of the universe shrinks by a very modest amount, while the Hubble rate increases tremendously. The ekpyrotic phase is followed by a brief phase of scalar kinetic energy domination, and during this phase the crunch/bang transition, which will be discussed in detail in the next section, occurs. Thus, the phases of expansion and contraction are highly asymmetric: in particular, over the course of one cycle, the universe grows by a huge amount. Only local quantities, such as the Hubble rate for example, behave cyclically, and return to the same value cycle after cycle.

During the ekpyrotic phase, as the scalar field rolls down the steep potential, quantum fluctuations are amplified and stretched beyond the horizon \cite{Khoury:2001zk}. This is because, although the scale factor remains almost constant, the Hubble rate grows rapidly or, equivalently, the Hubble horizon shrinks rapidly. Thus fluctuation modes of successively smaller wavelengths find themselves outside of the horizon. In this way, the ekpyrotic phase generates classical perturbations, similarly to inflation. However, it turns out that the fluctuations in the field $\phi$ driving the ekpyrotic phase have a blue spectrum, with spectral index $n_s \approx 3,$ and a small amplitude \cite{Khoury:2001zk,Lyth:2001pf,Tsujikawa:2002qc,Creminelli:2004jg}. Thus these perturbations are irrelevant on cosmological scales of interest.

To see how scale-invariant curvature perturbations are generated, it is instructive to first consider the embedding of the cyclic model in string theory. The higher-dimensional view of the cyclic universe is set in heterotic M-theory, which describes 11-dimensional supergravity compactified on a 6-dimensional Calabi-Yau space and with one spatial dimension being a line segment (orbifold) $S^1/\mathbb{Z}_2$ \cite{Lukas:1998yy,Lukas:1998tt}. The (10-dimensional) endpoints of the line segment are called orbifold branes. For the quantum consistency of the theory (\ie in order for anomalies to be absent), Ho\v{r}ava and Witten have shown that these branes need to have $E_8$ gauge fields that reside on their worldvolume \cite{Horava:1995qa,Horava:1996ma}. These gauge groups are large enough to easily contain the Standard Model of particle physics, and explicit realizations of the (supersymmetric) Standard Model have been realized in this context by adding supplementary branes in between the orbifold branes \cite{Braun:2005nv} (these supplementary branes do not concern us here). Thus, the 4-dimensional low energy effective theory of this setting can accommodate what we know about particle physics, while also containing additional scalar fields that can be relevant for cosmology. There are always at least two such scalars (moduli), one of which parameterizes the distance between the orbifold branes, and the other the volume of the Calabi-Yau space. Thus, from the higher-dimensional point of view, it is over-restrictive to consider a single scalar field.

As we will review in section \ref{SectionPhoenix}, the 4-dimensional effective action for these `universal' scalars is given by \be S=\int
\sqrt{-g}[R-\frac{1}{2}(\pt\phi_1)^2
-\frac{1}{2}(\pt\phi_2)^2-V(\phi_1,\phi_2)],\label{4dEFT}\ee  where
$\phi_1$ and $\phi_2$ are related by a field redefinition to
the inter-brane distance and the Calabi-Yau volume \cite{Lehners:2006ir}. If we now assume that both scalars obtain an ekpyrotic-type potential\footnote{No such potential has been explicitly derived from string theory to date, but there are indications that a potential of this form could arise due to membrane instantons stretching between the orbifold branes \cite{Lima:2001jc,Moore:2000fs}.},
 \be V(\phi_1,\phi_2) =-V_1 e^{-c_1 \phi_1} - V_2
e^{-c_2 \phi_2}, \label{potential2field}\ee where $V_1,V_2$ are constants and $c_1,c_2$ slowly varying functions of $\phi_1,\phi_2,$ then both the background dynamics and the characteristics of the generated perturbations change drastically. It turns out to be much
more natural to discuss the dynamics in terms of the rotated scalars $\s$ and $s$ pointing transverse and perpendicular to
the field velocity respectively
\cite{Koyama:2007mg,Koyama:2007ag}; they are defined, up to
unimportant additive constants which we will fix below, via \be
\s \equiv \frac{\dot\phi_1 \phi_1 + \dot\phi_2 \phi_2}{\dot\s},
\qquad s \equiv \frac{\dot\phi_1 \phi_2 - \dot\phi_2
\phi_1}{\dot\s}, \ee with $\dot\s \equiv (\dot\phi_1^2 +
\dot\phi_2^2)^{1/2}.$ It is also useful to define the angle
$\th$ of the trajectory in field space, via
\cite{Gordon:2000hv} \be \cos \th =\frac{\dot\phi_1}{\dot\s},
\qquad \sin \th = \frac{\dot\phi_2}{\dot\s}. \ee Then the potential can be re-expressed as \be
V=-V_0 e^{\sqrt{2\e}\s}[1+\e s^2+\frac{\k_3}{3!}\e^{3/2}
s^3+\frac{\k_4}{4!}\e^2 s^4+\cdots],
\label{potentialParameterized}\ee where for exact exponentials
of the form (\ref{potential2field}), one has
$\k_3=2\sqrt{2}(c_1^2-c_2^2)/|c_1 c_2|$ and $\k_4=4(c_1^6 +
c_2^6)/(c_1^2 c_2^2(c_1^2 + c_2^2)).$ Here $\epsilon,$ which is defined by $1/\e = 2c_1^2 + 2c_2^2$ and characterizes the steepness of the potential, is related to the equation of state $w$ via $\e = \frac{3}{2}(1+w),$ and is assumed to be slowly varying.
The potential describes an ekpyrotic direction along $\s$, combined with a transverse unstable direction along $s.$ This instability is key to what follows.
It implies that there is a ridge at $s=0,$ and the evolution along this ridge is characterized by the ekpyrotic scaling solution
\be a(t)=(-t)^{1/\e} \qquad
\s=-\sqrt{\frac{2}{\e}}\ln \left(-\sqrt{\e V_0} t\right) \qquad
s=0, \label{ScalingSolution}\ee with the angle $\th$ being
constant.

As it is $\s$ that drives the ekpyrotic phase, this scalar field develops fluctuations with a blue spectrum. However, as shown in \cite{Lehners:2007ac}, the transverse field $s$ obtains (isocurvature) fluctuations with an amplitude and spectral index given by
 \bea
Q_s &\approx& |\e V_{\mathrm{turn}}|^{1/2} \label{entropyamplitude}\\ \label{tilt1} n_s -1 &=& \frac{2}{\epsilon } -
\frac{\epsilon_{,N}}{\epsilon^2}, \eea
where the meaning of $V_{\mathrm{turn}}$ will be explained momentarily. If $\e \sim {\cal O}(10^2)$ or more, the spectrum is close to scale-invariant. In fact, given that the ekpyrotic phase must come to an end, the potential is expected to be less steep in the bottom half, leading to a natural range of $0.97 < n_s < 1.02$ \cite{Lehners:2007ac}.

But what we observe in the CMB are density fluctuations that are caused by fluctuations in the curvature of the universe. Thus, to match the model with data, we must calculate its predictions for the curvature perturbation $\z.$ On large scales, and at linear order, the curvature perturbation evolves according to \cite{Gordon:2000hv} \be \dot\z =
-\frac{2H}{\dot\s}\dot\th \d s = \sqrt{\frac{2}{\e}}\dot\th \d
s. \label{zetalinear}\ee Hence, if the trajectory in scalar field space turns ($\dot\th \neq 0$), then the isocurvature perturbations $\d s$ act as a source for the curvature perturbations $\z,$ which, given the absence of momentum-dependence in (\ref{zetalinear}), will acquire the identical spectrum \cite{Finelli:2002we,Notari:2002yc}. Such a conversion of isocurvature into curvature perturbations can occur in two ways:

\begin{itemize}
\item  {\it Ekpyrotic Conversion:} if the trajectory is not localized close enough to the ridge during the ekpyrotic phase, then the trajectory will turn at a value of the potential $V_{\mathrm{turn}},$ and subsequently roll down one of the steep sides of the potential\footnote{During this further rolling down, the universe is still in the ekpyrotic phase, albeit with a slightly different equation of state. However, now there is no further transverse instability, and the additional isocurvature perturbations that are produced during this stage of evolution have a blue spectrum and are irrelevant for cosmology, even if they should get converted into curvature perturbations at a later time due to a further turning of the trajectory.}\cite{Buchbinder:2007ad,Creminelli:2007aq,Koyama:2007mg}. The amplitude of the nearly scale-invariant curvature perturbations that are generated in this process can be estimated by combining (\ref{entropyamplitude}) and (\ref{zetalinear}), yielding \be Q_\z^2 \approx |V_{\mathrm{turn}}|.\ee Thus, the further down the potential the turn occurs, the more structure will be produced in the corresponding region of the universe. As shown in \cite{Buchbinder:2007at,Koyama:2007if}, this conversion mechanism also generates large local non-gaussian corrections: writing the full curvature perturbation as a linear, gaussian piece $\z_g$ plus correction terms as
\be \z = \z_g + \frac{3}{5}f_{NL} \z_g^2 + \frac{9}{25}g_{NL}\z_g^3, \ee
the predictions for the bi- and trispectrum are
\bea f_{NL} &=& -\frac{5}{12} c_1^2 \qquad \qquad \qquad {\text{Ekpyrotic Conversion}}\\
g_{NL} &=& \frac{25}{108}c_1^4,\eea where we have assumed that after the turn the trajectory evolves approximately in the $\phi_2$ direction (otherwise the formulae above should contain $c_2$ instead of $c_1$).

\item {\it Kinetic Conversion:} if the trajectory is localized close enough to the ridge during the ekpyrotic phase, so that the field makes it all the way down, then, as discussed in more detail in the next section, the trajectory automatically turns during the kinetic phase, when $\e \approx 3$ \cite{Lehners:2007ac}. In this case, the amplitude can be estimated again by combining (\ref{entropyamplitude}) and (\ref{zetalinear}) to be
    \be Q_\z^2 \approx \e_{\mathrm{ek}} |V_{\mathrm{min}}|,\ee where $V_{\mathrm{min}}$ stands for the potential value at the bottom of the ekpyrotic potential. Thus, compared to the maximal amplitude achieved by the ekpyrotic conversion mechanism, the amplitude $Q_\z$ is now enhanced by a factor of $\sqrt{\e_{\mathrm{ek}}}$ \cite{Lehners:2010fy}. Detailed calculations show that for realistic values, {\it e.g.} $\e_{\mathrm{ek}} \approx 50,$ the enhancement in $Q_\z$ is only a factor of about $2$, but nevertheless there is an enhancement. For this case of kinetic conversion, significant non-gaussian corrections are also produced, with detailed calculations yielding \cite{Lehners:2007wc,Lehners:2008my,Lehners:2009ja,Lehners:2009qu}
    \bea f_{NL} &=& \frac{3}{2}\k_3\, \sqrt{\e} + 5 \qquad \qquad {\text{Kinetic Conversion}}\label{fNLkinetic}\\ g_{NL} &=& (\frac{5}{3}\k_4+\frac{5}{4}\k_3^2-40)\, \e,\label{gNLkinetic}\eea where we may expect $\k_3,\k_4 \sim {\cal O}(1).$

\end{itemize}

Note that, as presented here and as conceived in the literature so far \cite{Koyama:2007mg,Buchbinder:2007tw}, the two mechanisms of conversion appear to be rather distinct, and appear to depend on the specifics of a given model, in particular on the initial conditions assumed or generated at the beginning of the ekpyrotic phase. As we will demonstrate now, in the cyclic model of the universe this is not so and in fact {\it all} field trajectories are simultaneously physically realized! To make this statement explicit, we must look in more detail at the brane collision and the associated bounce phase.

\section{Brane Collisions and the Phoenix Universe} \label{SectionPhoenix}

The central part of the cyclic universe is the bounce phase, where the transition from contraction to expansion occurs. In a flat universe, such a transition can only occur if the null energy condition is violated, and thus we need physics that goes beyond the ordinary 4-dimensional Standard Model of particle physics to describe such an event. Indeed, string theory contains many objects that violate the null energy condition (such as orientifolds, negative-tension orbifolds, anti-branes), but which nevertheless do not necessarily lead to inconsistencies regarding, say, the second law of thermodynamics. We will focus on the case where the bounce is caused by the collision of a positive- and a negative-tension orbifold plane in (heterotic) M-theory. Such a collision is singular from the 4-dimensional point of view, and thus the null energy condition is only violated at a single moment in time, and not over an extended period. There have also been attempts to model non-singular bounces \cite{Buchbinder:2007ad,Creminelli:2007aq,Creminelli:2010ba}, but it is not clear yet whether such descriptions are consistent or not \cite{Xue:2010ux,Xue:2011nw}, and whether such non-singular bounces can arise in string theory or not \cite{Adams:2006sv,Kallosh:2007ad,Khoury:2010gb,Khoury:2011da}. The (classically singular) collision of orbifold planes currently represents the most promising description of a bounce. This brane collision has been analyzed both classically and semi-classically in a number of works \cite{Turok:2004gb,Lehners:2006pu,Niz:2006ef,Niz:2007gq,Copeland:2010yr}, and we will base our study of the cyclic universe on these works.

It was shown by Ho\v{r}ava and Witten that the strongly coupled heterotic string theory can be envisioned as 11-dimensional supergravity with one spatial direction being a line segment, or orbifold $S_1/\mathbb{Z}_2$ \cite{Horava:1995qa}. The size of this orbifold dimension determines the string coupling constant (which is larger, the larger the orbifold), and the endpoints of the line segment describe two orbifold planes of opposite tensions. For the quantum consistency of the theory, \ie for the absence of anomalies, these two orbifold planes must contain gauge fields with symmetry group $E_8.$ In order to obtain a realistic model of particle physics, 6 of the spatial dimensions should be compact and in the shape of a Calabi-Yau manifold (in order to ensure minimal supersymmetry in the 4-dimensional effective theory). It turns out that, in order to reproduce the observed value of Newton's constant, one needs to consider configurations in which the Calabi-Yau radius is smaller by one or two orders of magnitude than the orbifold dimension \cite{Witten:1996mz}. Thus it makes more sense to dimensionally reduce the original 11-dimensional theory on a Calabi-Yau manifold first, and then investigate whether this theory has braneworld-type solutions, in which two branes are separated along an orbifold dimension. Such a dimensional reduction was performed in \cite{Lukas:1998yy}, with the metric being reduced according to \be ds_{11}^2 = V_{CY}^{-2/3}ds_5^2 + V_{CY}^{1/3}ds_{CY}^2, \label{11dmetric}\ee where $V_{CY}$ denotes the volume of the Calabi-Yau manifold. The resulting 5-dimensional effective theory, called heterotic M-theory, has action \bea S_5 &=&
\frac{1}{2\k_5^2}\int_{5d}\sqrt{-g}\left[
                  R-\frac{1}{2}V_{CY}^{-2}\partial_m V_{CY}\partial^m V_{CY}
                   -6 \a^2 V_{CY}^{-2}\right] \nn
                   \\ && +\frac{1}{2\k_5^2}\left\{-12\a \int_{4d,y=+1}\sqrt{-g}
                   \, V_{CY}^{-1} +12 \a \int_{4d,y=-1}\sqrt{-g}\,
                   V_{CY}^{-1} \right\} \; ,  \label{5dAction}
                   \eea
where we have kept only the cosmologically relevant fields, \ie gravity and the Calabi-Yau volume modulus $V_{CY}$ (the index $m$ runs over time and 4 spatial dimensions). Here $\a$ denotes the amount of 4-form flux wrapping a 4-cycle of the Calabi-Yau manifold, and must be non-zero in order to obtain braneworld solutions. In fact, as by assumption $\a \neq 0,$ the vacuum of this theory is not given by 5-dimensional Minkowski space, but rather by the braneworld spacetime \cite{Lukas:1998yy}\bea
d s^2 &=& h^{2/5}(y)\,\big[A^2 \,(-d t^2 + d \vec{x}^2) + B^2 \,d y^2\big], \nn \\
 V_{CY} &=& B\, h^{6/5}(y), \label{domainwall} \nn \\
h(y) &=& 5\a\, y+C,
\eea
where $A$, $B$ and $C$ are arbitrary constants.
The $y$ coordinate is taken to span
the orbifold $S^1/\mathbb{Z}_2$ with fixed points at $y=\pm 1$.
In an `extended' picture of the solution, obtained by $\mathbb{Z}_2$-reflecting the solution across the branes,
there is a downward-pointing kink at $y=-1$ (corresponding to a negative-tension brane) and an upward-pointing kink at $y=+1$ (corresponding to a positive-tension brane).
By inspection, one can see that the volume of the
Calabi-Yau manifold and the distance between the boundary branes
are determined in terms of the moduli $B$ and $C$, while the scale
factors on the branes are determined in terms of $A$ and $C$. In a cosmological context, we are interested in describing the motion (and eventual collision) of these boundary branes. In order to do so, we can employ the moduli space approximation, which is valid as long as the branes move slowly (this will turn out to be a consistent requirement). The moduli space approximation consists of letting the `constants' $A,B,C$ become functions of time (for convenience, we use a conformal time coordinate $\t$ here). These moduli describe the lightest fields in the theory (as shown in \cite{Lehners:2005su}, all other perturbations are massive) and the action specifying their dynamics can be obtained by plugging the now time-dependent moduli back in the action, yielding \cite{Lehners:2006ir}
\bea
S_{\mathrm{mod}} = -3 \int_{4d}
A^2BI_{\frac{3}{5}} && [ \Big(\frac{{A'}}{A}\Big)^2-\frac{1}{12}\Big(\frac{{B'}}{B}\Big)^2
 +\frac{{A'}{B'}}{AB} \nn \\ &&
-\frac{1}{25}\frac{I_{-\frac{7}{5}}}{I_{\frac{3}{5}}}\,{C'}^2
+\frac{3}{5}\,\frac{I_{-\frac{2}{5}}{A'}{C'}}{I_{\frac{3}{5}}A}],
\label{ActionMSA1}
\eea
where we have defined
\bea
I_n &=& \int_{-1}^{1} dy \ h^n = \frac{1}{5\a(n+1)}[(C+5\a)^{(n+1)}-(C-5\a)^{(n+1)}].
\eea
This action can be greatly simplified by
introducing the field redefinitions
\bea
a^2 &\equiv& A^2\,B\,I_{\frac{3}{5}}, \\
e^{\f_1/\sqrt{2}} &\equiv& B\,(I_{\frac{3}{5}})^{3/4}, \\
\f_2 &\equiv& -\frac{\sqrt{6}}{20}\int d C\, (I_{\frac{3}{5}})^{-1}\,
\big[9\,(I_{-\frac{2}{5}})^2+16\,I_{-\frac{7}{5}}I_{\frac{3}{5}}\big]^{1/2}.
\label{DefinitionChi}
\eea
The definition
(\ref{DefinitionChi}) can be inverted to give \be C = 5\a \left[
\frac{(1+e^{2\sqrt{2/3}\f_2})^{5/4}
+(1-e^{2\sqrt{2/3}\f_2})^{5/4}}{(1+e^{2\sqrt{2/3}\f_2})^{5/4}
-(1-e^{2\sqrt{2/3}\f_2})^{5/4}}\right]. \label{relCmoduli} \ee In terms of
$a$, $\f_1$ and $\f_2,$ the moduli space action (\ref{ActionMSA1})
then reduces to the remarkably simple form \cite{Lehners:2006ir} \be S_{\mathrm{mod}} =
 \int_{4d} [-3{a'}^2 + \frac{1}{2}a^2 (\f_1^{'2} + \f_2^{'2})].
\label{ActionMSA2} \ee The minus sign in front of the kinetic term
for $a$ is characteristic of gravity, and in fact this is the
action for gravity with scale factor $a$ and two minimally coupled
scalar fields that was used in the previous section, Eq. (\ref{4dEFT}).
The resulting equations of motion are solved by  \bea
a^2 &=& a_0^2\,(\t_0-\t), \label{MSASolutiona}\\
\phi_1 &=& -\sqrt{\frac{3}{2}}\cos\th\,\ln [\f_{1,0}(\t_0-\t)], \label{MSASolutionphi1}\\
\phi_2 &=& -\sqrt{\frac{3}{2}}\sin\th\,\ln [\f_{2,0}(\t_0-\t)], \label{MSASolutionphi2}
\eea
where $a_0$, $\th$, $\t_0$, $\f_{1,0}$, and
$\f_{2,0}$ are constants of integration.
Thus, in the absence of a potential, the solutions to the equations of motion correspond to straight line trajectories at angle $\th$ in the scalar field space spanned by $\phi_1$ and $\phi_2.$

\begin{figure}[t]
\begin{center}
\includegraphics[width=0.75\textwidth]{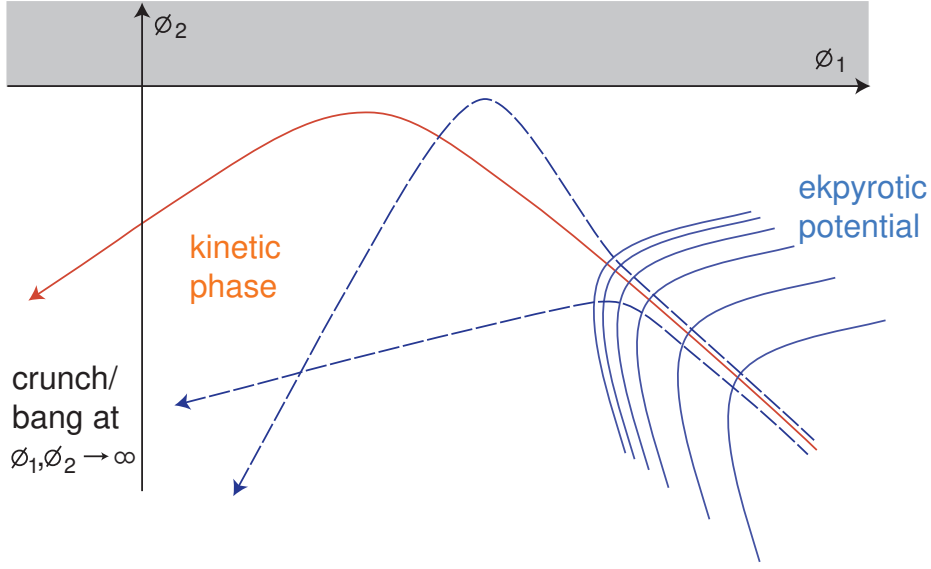}
\caption{\label{figureTrajectories} {\small Examples of trajectories at different angles in scalar field space. The solid line trajectory corresponds to the kinetic conversion mechanism, and the dashed trajectories to the ekpyrotic conversion mechanism.
}}
\end{center}
\end{figure}

This scalar field space has one peculiar feature, which is best understood by first using the metrics (\ref{11dmetric}) and (\ref{domainwall}) to relate 11-dimensional geometric quantities, in particular the inter-brane distance $d_{11}$ and the Calabi-Yau volume $V_{CY\pm}$ at the locations $y = \pm1$, to the 4-dimensional fields:
\bea
d_{11} &=& B^{2/3} I_{-1/5} \sim e^{\sqrt{2}\phi_1/3 + \sqrt{2}\phi_2/\sqrt{3}}, \label{relModDistance} \\
V_{CY\pm} &=& (2\a)^{3/4} \, e^{\f_1/\sqrt{2}} \,\left\{
\begin{array}{lll}
                     \left( \cosh \sqrt{2/3}\f_2 \right)^{3/2}  & \\
                     \left(-\sinh \sqrt{2/3}\f_2 \right)^{3/2} & \\
                    \end{array} \right. \sim e^{\phi_1/\sqrt{2} - \sqrt{3}\phi_2/\sqrt{2}}.  \label{relModCYvolume}
\eea
The approximate relations on the right hand side exhibit the dependence of $d_{11}$ and $V_{CY\pm}$ on the scalar fields as the branes become close, \ie as $\phi_2 \rightarrow - \infty.$ Eq. (\ref{relModCYvolume}) shows that the volume of the Calabi-Yau manifold at the location of the negative-tension boundary brane would become negative, if $\phi_2$ became positive. As shown in \cite{Lehners:2007nb}, this unphysical region is generally not accessible, and the axis $\phi_2=0$ acts as a repulsive boundary - in other words, there is an effective repulsive potential near this axis. Thus trajectories approaching the boundary bend in its vicinity, and this turning automatically causes the conversion of isocurvature perturbations into curvature perturbations, as used above in the discussion of the kinetic conversion mechanism - see also Fig. \ref{figureTrajectories} for an illustration.

Fig. \ref{figureTrajectories} depicts three different field trajectories, the solid line corresponding to a trajectory that makes it all the way down the ridge in the ekpyrotic potential and turns at the $\phi_2=0$ boundary before shooting off to $-\infty$ where the brane collision occurs. The dashed lines in the figure show two examples of trajectories that turn before reaching the bottom of the potential. These correspond to the ekpyrotic conversion mechanism. After their turn, they keep rolling down a steep side of the potential, and then shoot off to $-\infty$ as well, either directly, or after bouncing off the moduli space boundary at $\phi_2=0$ first. If such a second turn occurs, this will have little effect on the cosmological density perturbations, since the additional isocurvature modes that are generated after the first turn have a blue spectrum, and have wavelengths that are too small to be of interest. In fact, the currently observable modes are those generated between about 50 and 60 e-folds before the end of the ekpyrotic phase, \ie when the scalar field is a little more than halfway down the potential (assuming the potential minimum is approximately at the GUT scale). This means that for trajectories turning during the ekpyrotic phase, the turn must occur in the lower half of the potential if scale-invariant perturbations are to be generated on CMB scales. One may also wonder what happens to trajectories that turn early on. In order to conjecture about the fate of those trajectories, it is useful to re-consider the general requirements on the ekpyrotic potential: these are that it must be steep and negative, with a transverse unstable direction. It seems to be too much to assume that this transverse unstable direction will continue very far away from the ridge. It seems more reasonable to guess that trajectories that turn early will not undergo a long second ekpyrotic phase down the side of the potential, but rather that the ekpyrotic phase will come to a premature end. If this is so, the regions corresponding to those trajectories will develop strong curvature anisotropies in the approach to the brane collision, and will undergo a BKL-type mixmaster crunch \cite{Belinsky:1970ew}. We conjecture that these highly curved crunches will result in an over-production of particles at the brane collision, and that these regions will remain in a gravitationally collapsed state - locally, the branes will stick together. These regions stop growing and stop cycling.

It is worth studying those trajectories that approach the brane collision after a long ekpyrotic phase in more detail. The collision occurs as $\phi_1,\phi_2 \rightarrow - \infty,$ \ie the trajectories of interest are those described by Eqs. (\ref{MSASolutiona}) to (\ref{MSASolutionphi2}), with $\pi < \th < \frac{3\pi}{2}.$ We can choose $\t_0 = 0,$ so that the collision occurs as $\t \rightarrow 0^-.$ Using (\ref{11dmetric}) and (\ref{domainwall}), we can lift these solutions to 11 dimensions, where (in the $\phi_2 \rightarrow - \infty$ limit) the metric becomes
\bea
g_{\mu \nu} &\approx& \eta_{\mu \nu} (2\a)^{-1/4} a^2 e^{-5\phi_1/(3\sqrt{2})+\phi_2/\sqrt{6}} \sim (-\t)^{1+5\cos\th/(2\sqrt{3})- \sin\th/2} \\
g_{yy} &\approx& (8\a)^{-1} e^{4\phi_1/(3\sqrt{2})+2\sqrt{2}\phi_2/\sqrt{3}} \sim (-\t)^{-2\cos\th/\sqrt{3}-2 \sin\th} \\
g_{ab} &\approx& g^{CY}_{ab} (2\a)^{1/4} e^{\phi_1/(3\sqrt{2})-  \phi_2/\sqrt{6}} \sim (-\t)^{-\cos\th/(2\sqrt{3})+ \sin\th/2},
\eea
with the $a,b$ indices denoting the Calabi-Yau directions, and the intrinsic Calabi-Yau metric being denoted $g^{CY}_{ab}.$ Defining a new time variable $T$ via \be -\t \propto (-T)^{4/(6+5\cos\th/\sqrt{3}-\sin\th)},\ee the metric can be brought into Kasner form \be ds^2 = -dT^2 + \sum_{i=1,...,10} T^{2p_i} dx_i^2, \ee
where $i$ runs over all spatial dimensions. Then, denoting the exponents for the ordinary spatial dimensions by $p_1=p_2=p_3\equiv p_{3d},$ the one for the orbifold by $p_y$ and those for the Calabi-Yau directions by $p_5=p_6=p_7=p_8=p_9=p_{10}\equiv p_{CY},$ we get
\bea
p_{3d} &=& 1-\frac{4}{6+5\cos\th/\sqrt{3}-\sin\th}\\
p_y &=& \frac{-4\cos\th/\sqrt{3}-4\sin\th}{6+5\cos\th/\sqrt{3}-\sin\th} \\
p_{CY} &=& \frac{-\cos\th/\sqrt{3}+\sin\th}{6+5\cos\th/\sqrt{3}-\sin\th}.
\eea
These solutions correspond to 11-dimensional Kasner solutions, with $\sum p_i = \sum p_i^2 = 1$, but where the three ordinary spatial dimensions as well as the Calabi-Yau dimensions remain isotropic amongst themselves. In fact, as the orbifold dimension shrinks, the ordinary three spatial dimensions grow while the Calabi-Yau shrinks if $\pi < \th < \frac{7\pi}{6},$ and vice versa for $\frac{7\pi}{6} < \th < \frac{3\pi}{2}.$ The intermediate value $\th=\frac{7\pi}{6}$ corresponds to the special case where the spacetime is Milne, with both the ordinary and the Calabi-Yau spatial dimensions approaching a constant value at the collision.

The Milne solution was analyzed semi-classically in \cite{Turok:2004gb}, and the Kasner solutions more recently in \cite{Copeland:2010yr}. In all of these solutions, the curvature is small in the approach to the collision, and thus curvature corrections are suppressed. In fact, we may picture the effect of the preceding ekpyrotic phase as flattening and parallelizing the branes locally. As the branes become close, the lightest states in the theory correspond to M2-branes stretching between the two boundary branes. Their equations of motion remain regular at the collision, provided the Kasner solutions are ``close'' to Milne in the sense that $p_i + p_y/2 > 0$ for all $i.$ This condition means that no direction should expand faster than (the square root of) the scale factor of the orbifold shrinks, and it is straightforward to verify that this condition is satisfied for all solutions above. Moreover, using semi-classical methods, the amount of particle (\ie membrane) production at the collision was estimated and found to be proportional to the collision rapidity. For non-relativistic collision speeds, the amount of particle production is small, and we conjecture that universes undergoing such a bounce re-emerge unscathed and proceed to a new expanding phase. By contrast, for relativistic collision speeds, we may assume that there is so much particle production that the corresponding regions of the universe over-close and re-collapse rapidly. Such regions, along with those that did not undergo a prolonged ekpyrotic phase and enter the collision in a highly curved state, are filtered out by the brane collision, and effectively decouple from the expanding regions.

We have just conjectured that there is a process of dynamical selection occurring at the big crunch/big bang transition. What are its consequences? Consider a region of space that is at the end of its dark energy phase, and about to start ekpyrosis. As shown in \cite{Lehners:2008qe}, in scalar field space this region corresponds to a small blob sitting at the top of the ridge in the ekpyrotic potential, with a small spread $\Delta s $ of field values along the transverse direction and a negligible spread $\Delta \s$ along the ridge direction. Let $N_{\mathrm{turn}}$ denote the number of e-folds of ekpyrosis that occur before the trajectory turns. As discussed earlier, scale-invariant curvature perturbations on observable scales are only produced when this number is greater than about 70, and here we will assume this to be the case. Then a fraction $e^{-N_{\mathrm{turn}}}$ of trajectories will roll down the ridge up to the potential height $V_{\mathrm{turn}}$ where the trajectory will turn, thus producing curvature perturbations with amplitude $Q_\z \approx |V_{\mathrm{turn}}|^{1/2}$ and non-gaussian corrections of the ``ekpyrotic conversion'' type. Afterwards, the trajectory will continue along one of the steep sides of the potential until eventually the potential bottoms out, becomes unimportant, and a kinetic phase ensues. The trajectory will now shoot off to $-\infty$ in scalar field space, and there is no further instability along the rest of the trajectory. The brane collision will occur with a collision velocity $y_0$ determined by the minimum of the potential $V_{\mathrm{min}}$. As shown in \cite{Lehners:2010ug}, requiring this velocity to be non-relativistic implies a bound $|V_{\mathrm{min}}|\lesssim 10^{-10}$ in Planck units. For now, let us assume that this bound is saturated. This implies that rolling down the entire ridge in the potential corresponds to 120 e-folds of ekpyrosis. Thus, a fraction $e^{-120}$ of all trajectories undergo the full ekpyrotic phase and turn only during the kinetic phase. These ``kinetic conversion'' trajectories generate curvature perturbations with the largest amplitude, $Q_\z \approx |V_{\mathrm{min}}|^{1/2} \approx 10^{-4.5}$ \cite{Lehners:2010ug}, with the corresponding non-gaussian corrections, and they lead to a brane collision with approximately the same velocity $y_0.$ After the brane collision, all regions that made it through will expand by the huge factor of $e^{60 + N_{de}},$ as described in section \ref{SectionCyclic}. 

Above, we argued that it is unreasonable to assume that the turn in the trajectory can occur arbitrarily early during the ekpyrotic phase and still make it through the bounce. If the earliest that the turn can occur corresponds to $N_{\mathrm{turn,min}},$ then we must compare the losses occurring because of the instability of the ekpyrotic phase with the amplification occurring in the expanding phase, \ie we need $N_{\mathrm{turn,min}} < N_{de} + 60,$ or $N_{de}> N_{\mathrm{turn,min}} - 60,$ in order for the cycling to be sustainable. Thus dark energy remains crucial in ensuring the survival of the cyclic universe, but since $N_{\mathrm{turn,min}} \leq 120$, the constraint on the total amount of dark energy expansion is weakened from the earlier bound, which was $N_{de} > 60.$

\begin{figure}[t]
\begin{center}
\includegraphics[width=0.75\textwidth]{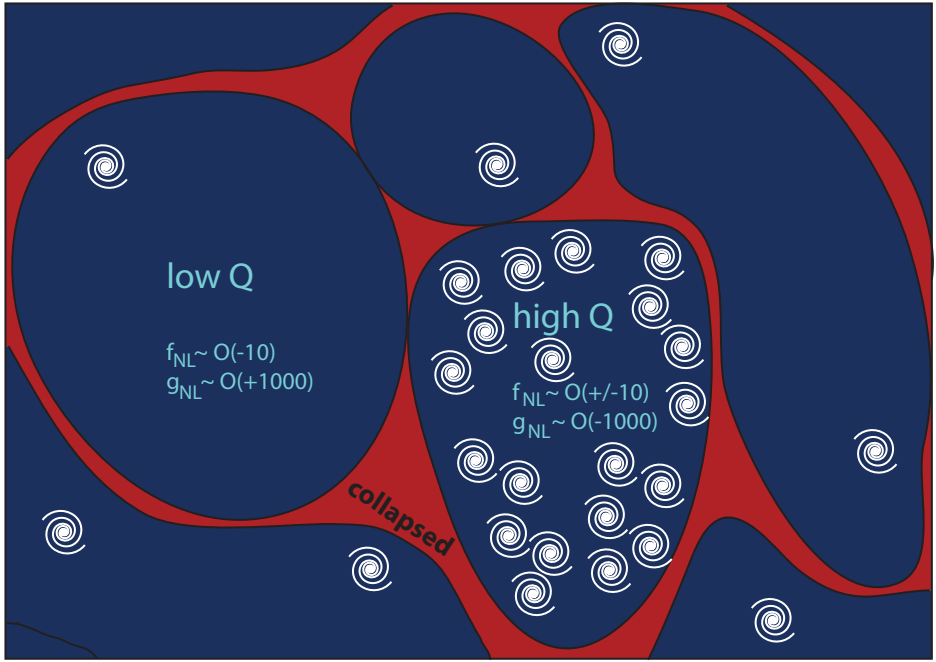}
\caption{\label{figurePhoenix} {\small A spatial slice through the phoenix universe.
}}
\end{center}
\end{figure}

The resulting global structure is an extension of the phoenix universe of \cite{Lehners:2008qe}, see Fig. \ref{figurePhoenix}: during the ekpyrotic phase, most field trajectories turn too early and end their ekpyrotic flattening phase prematurely. These regions end up in gravitationally collapsed structures, {\it e.g.} black holes. But some field trajectories, approximately $e^{-N_{\mathrm{turn,min}}}$ in number, undergo a long ekpyrotic phase, generate nearly scale-invariant curvature perturbations and make it through the bounce. These flat and potentially habitable regions get amplified hugely during the phases of radiation, matter and dark energy expansion, and they eventually spawn new cycles of evolution. In previous work \cite{Lehners:2008qe}, there were two assumptions that have now been relaxed: the first was that only the Milne bounce, \ie only the special trajectory at angle $\th = \frac{7\pi}{6},$ was viable. And the second was that this trajectory was assumed to correspond to the ``kinetic conversion'' mechanism. Now we see that both assumptions are unnecessary: the orientation of the ridge can be at any angle in scalar field space, and all trajectories that experience sufficient ekpyrosis make it through the bounce, as long as their angle is in the range $\pi < \th < \frac{3\pi}{2}.$ The consequence is that we now get a diverse structure in the phoenix universe, with a range of physical properties. The main segregation is still between large, flat regions and tiny, highly curved ones. But within the large, flat regions there is now considerable variety.

Of the flat regions, the rarest are those that make it all the way down the ekpyrotic potential and convert their isocurvature into curvature perturbations during the kinetic phase - but these regions also have the largest perturbation amplitude. By contrast, the regions that correspond to earlier turns of the trajectory will be far more numerous, but will also contain a lot less structure. Hence it is of interest to estimate the amount of structure formed in these various large regions. The full ekpyrotic phase corresponds to about \be N_{\mathrm{ek}} = \frac{1}{2}\ln \frac{|V_{\mathrm{min}}|}{V_0} \approx 120 \ee e-folds, where $V_0$ denotes the present energy density of dark energy. The regions that turn at $V_{\mathrm{turn}}$ undergo only \be N_{\mathrm{turn}} = \frac{1}{2}\ln \frac{|V_{\mathrm{turn}}|}{V_0}\ee e-folds of ekpyrosis. These regions are thus more numerous by a factor $e^{N_{\mathrm{ek}}-N_{\mathrm{turn}}},$ or, correspondingly, they occupy a volume that is larger by a factor of \be e^{3(N_{\mathrm{ek}}-N_{\mathrm{turn}})} = (\frac{V_{\mathrm{min}}}{V_{\mathrm{turn}}})^{3/2}.\ee The maximal density perturbation amplitude is achieved by the kinetic conversion regions - we will denote this amplitude $Q_{\mathrm{max}}.$ As discussed at the end of section \ref{SectionCyclic}, for realistic values of the equation of state, this amplitude is about twice as large as the largest amplitude that can be achieved with the ekpyrotic conversion mechanism. Given that $Q \propto |V|^{1/2},$ this implies that for a trajectory that turns at $V_{\mathrm{turn}},$ the associated density perturbation amplitude is \be Q_{\mathrm{turn}} \approx \frac{Q_{\mathrm{max}}}{2}(\frac{V_{\mathrm{turn}}}{V_{\mathrm{min}}})^{1/2}.\ee If we assume the Press-Schechter model of structure formation \cite{Press:1973iz}, then in a given region of space, the probability for having a fluctuation with amplitude $\d$ is proportional to $\frac{1}{Q} e^{-\d^2/Q^2}.$ Hence, the number of fluctuations of amplitude $\d$ in the regions turning at $V_{\mathrm{turn}},$ compared to the kinetic conversion regions, is
\be \frac{(\frac{V_{\mathrm{min}}}{V_{\mathrm{turn}}})^{3/2} \frac{1}{Q_{\mathrm{turn}}} e^{-\d^2/Q_{\mathrm{turn}}^2}}{\frac{1}{Q_{\mathrm{max}}} e^{-\d^2/Q_{\mathrm{max}}^2}} = 2 (\frac{V_{\mathrm{min}}}{V_{\mathrm{turn}}})^2 e^{-\frac{\d^2}{Q_{\mathrm{max}}^2}(4\frac{V_{\mathrm{min}}}{V_{\mathrm{turn}}}-1)}.\ee
The exponential suppression dominates over the power-law increase in volume, showing that there are far fewer large fluctuations in the ``ekpyrotic conversion'' regions. In fact, one can integrate over $V_{\mathrm{turn}}$ to estimate the number of fluctuations of amplitude $\d \sim Q_{\mathrm{max}}$ in all ekpyrotic conversion regions compared to the kinetic conversion ones, obtaining (with $dx \equiv dV_{\mathrm{turn}}/V_{\mathrm{min}}$)
\be \int_0^{V_{\mathrm{min}}} 2 (\frac{V_{\mathrm{min}}}{V_{\mathrm{turn}}})^2 e^{-(4\frac{V_{\mathrm{min}}}{V_{\mathrm{turn}}}-1)}\frac{dV_{\mathrm{turn}}}{V_{\mathrm{min}}} = \int_0^1 \frac{2}{x^2}e^{1-4/x}dx \approx 0.025.\ee
Hence, there are about 40 times as many density fluctuations of magnitude $10^{-5}$ in the kinetic conversion regions than in all other regions combined!

\section{Discussion}

The best understood version of the cyclic universe involves an ekpyrotic phase during which the potential is unstable. Because of this instability, nearly scale-invariant isocurvature perturbations are generated, which get converted into curvature perturbations as the trajectory in field space turns. This turn can happen in essentially two ways, either during the ekpyrotic phase or during the subsequent kinetic phase. It was thought up to now that these possibilities correspond to different cosmological models, depending on how the initial conditions are arranged. Here, we have shown that in the cyclic universe, where each cycle generates the ``initial conditions'' for the next one, all of these possibilities are physically realized in parallel.

This leads to a global structure of the phoenix type, where large, flat regions now come in a variety of physical characteristics, but, interestingly, with strong correlations between a number of cosmological observables. The regions containing by far the most structure have a primordial density fluctuation amplitude of $Q_{\mathrm{max}} \approx 10^{-4.5},$ with local non-gaussian corrections specified by $f_{NL} \sim {\cal O}(\pm 10)$ and $g_{NL} \sim {\cal O}(-1000).$ There are far more numerous regions with a continuous range of ever smaller density perturbation amplitudes (starting at about $Q_{\mathrm{max}}/2$) and local non-gaussianity parameters $f_{NL} \sim {\cal O}(- 10)$ and $g_{NL} \sim {\cal O}(+1000).$ Even when counted together, these regions contain far fewer large amplitude fluctuations though than the regions having the maximal possible density fluctuation amplitude, and thus most galaxies are to be found in the $Q_{\mathrm{max}}$ regions. Also, as is typical in cyclic models, none of these regions are expected to contain large amplitude primordial gravitational waves. Finally, there are regions that are in a gravitationally collapsed state. Those are regions where the ekpyrotic phase ended prematurely, and which (we conjecture) underwent a high-curvature brane collision from which they emerged with an overproduction of matter, causing them to re-collapse rapidly.

One can summarize the essential steps that lead to the global phoenix structure as follows: diversify, filter, amplify. The unstable potential diversifies the possible physical properties that the model produces, the brane collision filters out those regions that are sufficiently flat, and the radiation, matter and dark energy phases amplify those regions that made it through the bounce in good shape. This sequence of steps contributes to making the model highly predictive. Another aspect enhancing its predictivity is the fact that during the longest smoothing phase, \ie during the dark energy phase, the Hubble rate is lower than during the principal phases of structure formation (\ie the ekpyrotic, radiation and matter phases). This implies that even large quantum fluctuations during the dark energy phase do not change the structure of the model in an essential way (in contrast to eternal inflation). The only consequence of such fluctuations are that the corresponding regions expand a little more, and enter their next cycle a little later than the surrounding regions. A more quantitative analysis of this particular point will be left for future work.

From a model-building perspective, the new phoenix universe presented here has two advantages over the ``old'' version. First, it shows that no fine-tuning is needed regarding the orientation of the unstable ridge of the ekpyrotic potential. And secondly, the constraint on the number of e-folds of dark energy needed to sustain the cyclic universe is weakened. In fact, in the new picture, it is likely that our predecessor-universe was a region with low $Q$ and very little structure. 

In this framework, there are many opportunities for further study: one may wonder for example how sharp the boundaries between the different regions in the phoenix universe are, and whether these boundaries lead to any interesting effects. Evidently, it would be desirable to have a fully quantum calculation of the bounce, but such a treatment may still be quite far into the future. In the meantime, it would certainly be worthwhile to study more detailed models of the bounce, and perhaps investigate under what circumstances physical quantities can change. An interesting possibility is that the bounce may be a strong filter for many other physical quantities too, and thus may select a whole range of features of our universe that otherwise might appear to be merely historical accidents.

\section*{Acknowledgments}

I would like to thank Adam Brown, Matt Johnson, Justin Khoury, Paul Steinhardt, Neil Turok and David Wands for stimulating discussions, and the Laboratoire APC at Paris 7/Diderot as well as the Perimeter Institute for Theoretical Physics for their outstanding hospitality. The author gratefully acknowledges the European Research Council's support in the form of a Starting Grant.

\bibliography{KasnerPhoenix}

\end{document}